# Astro2020 APC Whitepaper

# High-Contrast Testbeds for Future Space-Based Direct Imaging Exoplanet Missions

**Principal Author:**
Name: Johan Mazoyer
Institution: Jet Propulsion Laboratory/California Institute of Technology/NASA Hubble Fellow
Email: johan.mazoyer@jpl.nasa.gov
Phone: (818) 393-3602

**Thematic Areas:** State of the Profession, Exoplanets, Optical Instrumentation

**Co-Authors:** Pierre Baudoz (Observatoire de Paris), Ruslan Belikov (NASA Ames), Brendan Crill (JPL/Caltech), Kevin Fogarty (Caltech), Raphaël Galicher (Observatoire de Paris), Tyler Groff (GSFC), Olivier Guyon (Subaru Telescope), Roser Juanola-Parramon (NASA GSFC), Jeremy Kasdin (Princeton), Lucie Leboulleux (Observatoire de Paris), Jorge Llop Sayson (Caltech), Dimitri Mawet (Caltech), Camilo Mejia Prada (JPL/Caltech), Bertrand Mennesson (JPL/Caltech), Mamadou N'Diaye (Observatoire de la Côte d'Azur), Marshall Perrin (STScI), Laurent Pueyo (STScI), Aki Roberge (NASA GSFC), Garreth Ruane (JPL/Caltech), Eugene Serabyn (JPL/Caltech), Stuart Shaklan (JPL/Caltech), Nicholas Siegler (JPL/Caltech), Dan Sirbu (NASA Ames), Rémi Soummer (STScI), Chris Stark (STScI), John Trauger (JPL/Caltech), Neil Zimmerman (NASA GSFC)

**Abstract:** *Instrumentation techniques in the field of direct imaging of exoplanets have greatly advanced over the last two decades. Two of the four NASA-commissioned large concept studies involve a high-contrast instrument for the imaging and spectral characterization of exo-Earths from space: LUVOIR and HabEx. This whitepaper describes the status of 8 optical testbeds in the US and France currently in operation to experimentally validate the necessary technologies to image exo-Earths from space. They explore two complementary axes of research: (i) coronagraph designs and manufacturing and (ii) active wavefront correction methods and technologies. Several instrument architectures are currently being analyzed in parallel to provide more degrees of freedom for designing the future coronagraphic instruments. The necessary level of performance has already been demonstrated in-laboratory for clear off-axis telescopes (HabEx-like) and important efforts are currently in development to reproduce this accomplishment on segmented and/or on-axis telescopes (LUVOIR-like) over the next two years.*

# 1. Key Science Goals and Objectives

Instrumentation techniques in the field of direct imaging of exoplanets have greatly advanced over the last two decades, with the development of stellar coronagraphy, in parallel with specific methods of wavefront sensing and control (WFS&C). **These studies have shown that we can now build a space-based instrument capable of detecting and spectral characterize Earth-like planets around Sun-like stars** (Stark et al. Astro2020 Science Whitepaper). Observations of our neighboring stars with such an instrument would allow us to determine if our solar system is unique or common, and would provide the possibility of assessing the habitability of exoplanets (Arney et al. and Robinson et al., Astro2020 Science Whitepapers). This has long been the goal of the exoplanet community, as shown by the National Academies 2019 reports on Exoplanet Science Strategy and Astrobiology Strategy for the Search for Life in the Universe, largely endorsed by our community (Plavchan et al. Astro2020 Science Whitepaper). NASA commissioned four large concept studies that represent possible future flagship missions that Astro2020 may consider to be of high scientific value to the astronomy and astrophysics community. Two of these projects involve a high-contrast instrument for the imaging and spectral characterization of exoplanets from space: LUVOIR (Luvoir Team 2018) and HabEx (Gaudi et al. 2018).

**This whitepaper describes the status of the experimental validation of the concepts currently being developed to build the coronagraphic instrument for these missions. It describes the goal and recent results of 8 optical high-contrast testbeds in the US and France**. It is complementary to two other APC Whitepapers presenting the latest developments in coronagraph designs (Shaklan et al. APC Whitepaper) in numerical simulation and in active WFS&C techniques (Pueyo et al. APC Whitepaper).

# 2. Technical Overview

The goal of a coronagraphic system is to remove a star's light to allow the imaging and characterization of its faint circumstellar companion. The flux ratio between an exo-Earth and a Sun-like star around which it orbits is $10^{-10}$ in the visible ($10^{-8}$ for Jupiters), at a few tens of mas.

## 2.1. Performance metrics

Performance metrics are usually normalized by the telescope diameter or collecting surface to be independent of the mission. The most widely metric used to evaluate the performance of a coronagraphic system is **raw contrast**, measuring, after the coronagraph and at a given separation, the faction of star light that was rejected by the system. However, we now know that a coronagraph system's performance cannot only be reduced to this metric (Ruane et al. 2018a) and further considerations affect coronagraphic performance as well. We must not only maximize starlight suppression but also minimize the effects of the instrument on the planet transmission and image (planetary throughput). The region over which the system is designed to provide high sensitivity to planets is called the dark-hole. The **inner working angle** (IWA) and **outer working angle** are the inner and outer radius of this dark-hole, usually expressed in elements of resolution of the telescope (λ/D, where D is the telescope diameter and λ the central wavelength). The **robustness** of a system evaluates its capacity to maintain its performance in presence of realistic dynamic aberrations created by the telescope in space (e.g. pointing errors). Finally, because the ultimate goal of coronagraphy is the spectral characterization of the detected exoplanets, an important parameter is the **spectral bandwidth** Δλ/λ for which we can achieve this performance, usually expressed in percentages.

Yield studies (Stark et al. 2019 and Science Whitepaper) have shown that the number of exo-Earths detected or characterized by a coronagraphic mission is a weak function of each of these individual parameters (contrast, planetary throughput, robustness, bandwidth, IWA & OWA). **Designing a coronagraph instrument is therefore a multi-variable optimization problem, constrained by the telescope geometry and stability. For this reason, it is crucial that very diverse solutions, each with their own advantages, be explored.**

### 2.2. Designing coronagraphic instruments for complex apertures

Coronagraph designs for clear aperture (HabEx-like aperture, off-Axis non-segmented, Fig. A1) were developed over the preceding two decades and are already widely used on ground-based telescopes. In the wake of WFIRST coronagraphic instrument (WFIRST/CGI) development, specific designs have been developed to obtain good performance with more complex apertures, such as off-axis missions (secondary mirror is not in the light path) with segmented primaries (LUVOIR-B architecture, Fig. A2). Finally, the central obscuration and spiders created by the secondary mirror in on-axis telescopes diffract light, degrading performance for monolithic telescopes (WFIRST, Fig. A3) or segmented telescopes (LUVOIR-A architecture, Fig. A4). Coronagraphs for these geometries have been developed in particular in the Segmented Coronagraph Design and Analysis (SCDA, Shaklan et al. APC Whitepaper) program, commissioned by NASA's Exoplanet Exploration Program (ExEP).

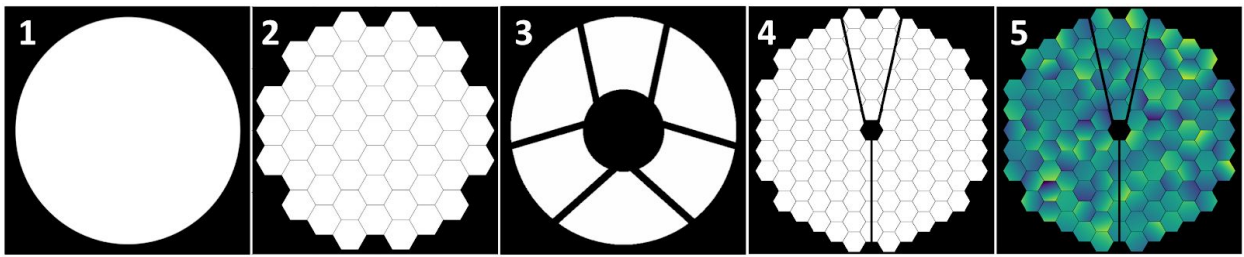

**Fig. A:** *Telescope apertures of increasing complexity. 1) Off-Axis non-segmented (clear aperture, **HabEx**); 2) Off-Axis segmented (**LUVOIR B**); 3) On-Axis non-segmented (**WFIRST**); 4) On-Axis segmented (**LUVOIR A**); 5) On-Axis segmented with segment misalignments*

But even the best designs are limited by the unknowns of the system (e.g. surface quality of the optics) and by dynamic aberrations created by thermal and mechanical evolution of the telescope (e.g. segments misalignment, Fig. A5) and pointing errors, which creates star light leaks in the science image. For this reason, **the success of a high-contrast testbed and by extension a high-contrast instrument cannot be reduced to its coronagraph design. For these complex systems, a comprehensive approach must simultaneously optimize three main leverages:**

  (i)   **A stable environment minimizing thermal variations and mechanical vibrations**
  (ii)  **A high contrast, high planetary throughput and robust coronagraph design**
  (iii) **A fast and stable active correction method with deformable mirrors (DMs) to correct for unknown or dynamic aberrations.**

The first point is a mechanical problem that cannot be simulated by the miniature modelling on an optical testbed. However, testbeds are simulating realistic space-based dynamic aberrations to show that coronagraph designs (ii) associated with WFS&C systems (iii) can obtain high performance in these conditions. The experimental validation of these complex systems is the purpose of this whitepaper.

## 3. Technology Drivers

The annually-updated list in ExEP's Technology Plan Appendix (Crill & Siegler 2019) defines the technology gaps and quantify, when possible, the difference between expected performance requirements and current state-of-the-art. Among those, we recall the two gaps specifically relevant to this whitepaper: coronagraph design and manufacture for clear or complex apertures and WFS&C algorithms and technologies to obtain and maintain high-contrast levels.

### 3.1. Coronagraph designs and manufacturing

Over the last two decades, four major different coronagraph techniques have emerged, combining apodization (change in the shape of the pupil or in its transmission to alter the associated diffraction pattern) and focal plane mask (localized amplitude and/or phase pattern in the focal plane to block and or diffract away the star light). They are now actively studied in the context of HabEx / LUVOIR: Hybrid Lyot Coronagraphs (HLC), Apodized Pupil Lyot Coronagraphs (APLC), Apodized Vortex Coronagraphs (AVC) and Phase Induced Amplitude Apodization Complex Mask Coronagraph (PIAACMC). **The SCDA study (Shaklan et al. APC Whitepaper) showed in numerical simulation that all these designs can reach the contrast level necessary to image Earth-like planets with complex apertures.** Four Strategic Astrophysics Technology (SAT) Technology Development for Exoplanet Missions (TDEM, see complete list) were awarded to study their design, manufacture, and experimental demonstration.

The *Super Lyot ExoEarth Coronagraph* TDEM (PI: J Trauger) studies HLC technology (Trauger et al. 2018). HLCs were selected for the WFIRST/CGI (Sec 4.2.1), but the performance in contrast has to be increased by a factor of 10 for HabEx / LUVOIR. In particular, the technologies to simultaneously manipulate the amplitude and phase with a metallic and a dielectric layer focal plane mask will be studied. The experimental validation takes place at JPL.

The *First System-level Demonstration of High-Contrast for Future Segmented Space Telescopes* (PI: Rémi Soummer) TDEM studies APLCs (Soummer 2019). This TDEM includes demonstration of carbon nanotubes technologies to create very efficient binary apodizers (Fig. C1) and testbed experimentation in coordination with WFS&C strategies for segmented apertures (e.g. control of segment misalignments in closed loop), first at STScI (Sec 4.1.3) and then at JPL (Sec 4.2.3). APLC coronagraph designs have been selected as the baseline for the LUVOIR A architecture (Fig. A4-5, Pueyo et al. 2017).

The *Vortex Coronagraph High-Contrast Demonstrations* (PI: E. Serabyn) TDEM is oriented towards developing the AVC for both off axis monolithic and segmented aperture (Serabyn et al. 2016). Vortex coronagraphs are the baseline design for the HabEx and LUVOIR B concepts. In particular, this TDEM will advance the technologies to demonstrate phase-only focal plane masks usable for exoplanet observations. The experimental validation will take place on testbeds located at Caltech (Sec 4.1.2) and JPL (Sec 4.2.2 and 4.2.3).

Finally the *Laboratory Demonstration of High-Contrast Using PIAACMC on a Segmented Aperture* (PI: R. Belikov) TDEM is studying PIAACMC which combines fixed-mirror based apodization and a focal plane mask to achieve performance at small IWA. Specifically, they are advancing high precision fabrication of complex mirror shape as well as photolithography and etching processes to produce the complex masks (Belikov et al. 2018a).

### 3.2. Active wavefront sensing and control methods and technologies

Active wavefront control using DMs are a key component of high-contrast imaging instrument (Pueyo et al. APC Whitepaper). To obtain a $10^{-10}$ dark-hole in the visible after a coronagraph, one must control the incoming light wavefront at a 10 pm level, which cannot be achieved using fixed systems only. **Active correction allows a fixed coronagraph design to be adapted to the reality of the high-contrast instrument, by correcting all of the unknown and/or dynamic aberrations in the wavefront path** (aberrations on the optics surface, thermal drift, evolution of the aperture). DMs can also be used in combination with coronagraphs to correct for aperture discontinuities directly (Mazoyer et al. 2018). One DM is enough to achieve high-contrast on a half (180°) dark-hole, but for the last decade, testbeds have used two sequential DMs to obtain better performance over symmetrical (360°) dark-holes (Pueyo et al. 2011).

Several WFS&C techniques have been developed for more than a decade, in particular on the High-Contrast Imaging Laboratory ([HCIL](), Princeton) testbed. Most facilities are using the Pair-Wise Probing technique for sensing and Electrical Field Conjugation (EFC) for control (Give'on et al. 2007). However, other WFS&C techniques are currently being investigated at Paris Observatory (Sec. 4.1.4) and Caltech (Sec. 4.1.2). Finally, a TDEM (PI: Olivier Guyon) was awarded to study linear dark field control (LDFC), designed specifically to maintain high-contrast dark-holes during a science sequence (Guyon et al. 2019), to be tested mainly in [University of Arizona WFS&C optical testbed](), Ames and Subaru (Sec 4.1.1)

**For high-contrast applications, the required specifications for the DMs are different than for classical ground-based adaptive optics (see review in Madec 2012). The actuator maximum strokes or speed are less important than minimum stroke, stability over time, surface quality, size, and number of actuators.** Two types of DMs, which offer a large number of actuators (up to 64x64 as of now), are currently being used on high-contrast testbeds:
- Microelectromechanical systems (MEMS) DMs made by Boston Micromachines are currently used on most high-contrast testbeds (see review Morgan et al. 2019).
- PMN (lead magnesium niobate) DMs from Xinetics have been used for more than a decade at JPL high-contrast experiments (Ealey & Trauger 2004).

Finally, beginning in 2019, a new spectroscopy-oriented technology effort began at Goddard Space Flight Center with the goal of maturing and demonstrating three subsystems for high-contrast systems: deformable mirrors with powered surfaces (Groff et al. 2016), novel integral field spectrograph designs, and hole-multiplying CCD detectors.

## 4. Organization, Partnerships, and Current Status:

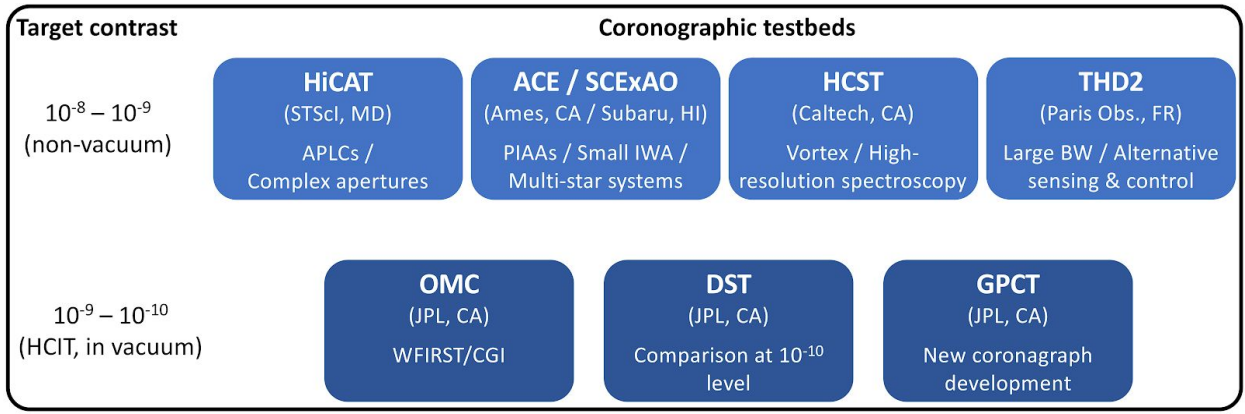

Diverse active WFS&C methods and coronagraphic designs, each with their own advantages, have to be studied. A coordinated effort has started to advance the validation process in parallel, with each team developing a specific expertise. In this section, we describe 8 high-contrast testbeds whose goal is to experimentally test both coronagraphic design and wavefront control methods specifically for space-based applications. Testbed performance is currently limited by the quality of its controlled environment (most importantly, vacuum chamber) and the components used (most importantly, the DMs). For this reason, performance comparisons between testbeds are not representative of the performances of the methods they are testing. Therefore, a fair comparison of two different coronagraph designs or active techniques has to be done in the same testbed facility and conditions, which is ultimately the goal of the Decadal Survey Testbed, which aims to achieve contrasts at the 10-10 level (Sec 4.2.3).

## 4.1 Non-vacuum testbeds

The testbeds in this section are not located in vacuum and are using MEMS DMs for correction of aberrations. For these reasons, their contrast performance is limited to a few $10^{-9}$ at best. To limit air variations, testbeds are placed inside enclosures in temperature-controlled clean rooms.

### 4.1.1 Ames Coronagraph Experiment (ACE) testbed

The Ames Coronagraph Experiment (ACE) is a state-of-the art facility located at NASA Ames and operated by a team led by Dr. Ruslan Belikov. **This testbed is focused on advancing fixed mirror apodization (PIAACMC) to provide high-contrast at small separations, enabling aggressive performance on relatively small telescopes**. It currently uses a single DM, a clear aperture and has the ability to simulate a binary star source. They have obtained a raw contrast of $2.10^{-8}$ from 2.4 - 4 λ/D in narrow bandwidth (Belikov et al. 2012).

More recently, the ACE team has been developing new WFS&C techniques. They have developed Multi-Star Wavefront Control (MSWC, Belikov et al. 2018b, Belikov et al. 2019 Science whitepaper) which allows the imaging of exoplanets in multi-star systems with existing mission concepts. MSWC increases the science yield of a high-contrast mission, including WFIRST, LUVOIR/HabEx, without any hardware modifications. Multi-star dark holes have been obtained experimentally. Finally, LDFC will be tested on this testbed to maintain already obtained high-contrast dark-holes during long period of time.

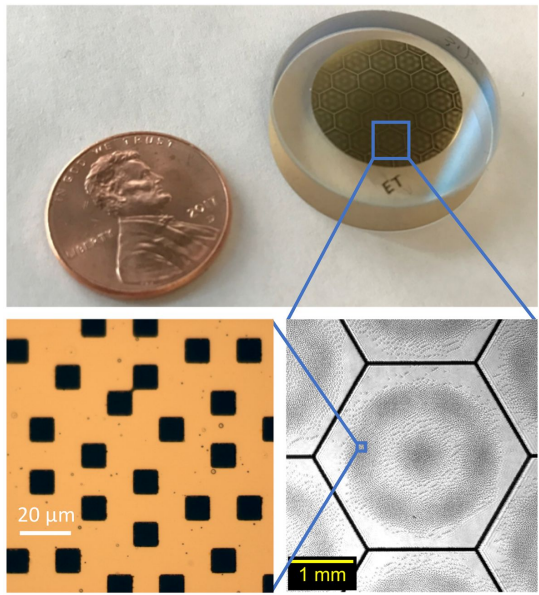

**Fig. B**: *HCST: a prototype apodizer for a vortex coronagraph on a LUVOIR-B (Fig A2) type telescope. The grayscale apodization is achieved with a binary pattern of 10 μm microdots.*

For PIAACMC and WFS&C validation, the ACE team is working in collaboration with the Subaru Coronagraphic Extreme Adaptive Optics (SCExAO), as part of the same TDEMs (LDFC and MSWC). This is a development platform for high contrast imaging techniques at the 8.2m Subaru Telescope, led by Dr. Olivier Guyon. **SCExAO is both a scientific instrument (night time) and a test platform. Its most valuable contribution is to provide a path to on-sky validation for advanced WFS&C techniques at moderate contrast levels.**

### 4.1.2 High-Contrast High-Resolution Spectroscopy for Segmented Telescopes (HCST)

The High-Contrast High- Resolution Spectroscopy for Segmented telescopes Testbed (HCST) is located at the California Institute of California (Pasadena, CA) and is operated by a team led by Dr. Dimitri Mawet. **They specialize in the use of Vortex coronagraphs (Ruane et al. 2018b), combining high-order phase mask with numerically optimized gray-scale apodizers** (Fig. B) specifically designed for off-axis segmented telescopes (LUVOIR B design, Fig A2), funded by the Vortex TDEM. This testbed is currently using a clear aperture and a single Boston DM, with plans to add a second one by the end of 2019. They recently started testing a new control algorithm (System ID, Sun et al. 2018), originally developed on the HCIL (Princeton, NJ). System ID adapts the linear model from data to include initially unknown effects and produce faster and more stable control.

Finally, **an important goal of this testbed is to explore solutions linking a classical coronagraph instrument to a spectrograph via a single mode optical fiber**. The fiber injection unit (FIU) was already demonstrated independently on a simpler version of the testbed (Llop Sayson, et al. 2019) and will be added to the testbed coronagraphic in July 2019. It will be used with a modified version of EFC to increase contrast.

### 4.1.3 High-contrast imager for Complex Aperture Telescope (HiCAT)

High-contrast imager for Complex Aperture Telescope (HiCAT) is a testbed located at the Space Telescope Science Institute and led by Dr. Rémi Soummer. The goal of the testbed is to study the impact of on-axis apertures (central obscuration, secondary struts, segmentation of the primary) on high-contrast imaging. The team uses a segmented mirror with 37 segments that can be controlled in piston, tip, and tilt to simulate the segmentation of the primary mirror and two serial MEMs DMs, currently controlled using a stroke minimization algorithm (Pueyo et al. 2011). The HiCAT team specializes in the use of APLCs, combining an apodizer (Fig C1), a focal plane mask, and a classical Lyot Stop, to correct for these discontinuities. They recently obtained their first dark-holes for clear aperture (Fig C3) and on-axis segmented aperture (Fig C2).

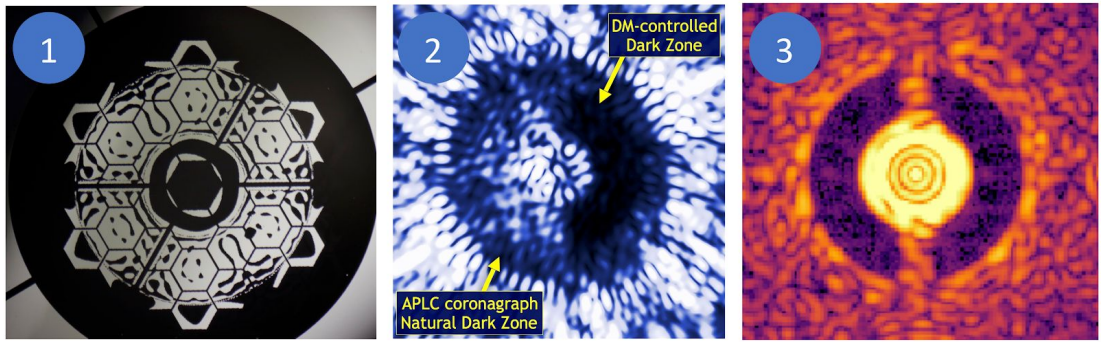

**Fig. C:** *HiCAT results. 1) apodizer for an on axis segmented aperture (LUVOIR A like). 2) dark-hole created with this apodizer in addition with DM. 3) Best current dark-hole with clear aperture.*

### 4.1.4 The Très Haute Dynamique 2 (THD2) testbed

The Très Haute Dynamique 2 (THD2) —French for "Very High-Contrast"— testbed is located at Paris Observatory (Meudon, France) and is operated by a team led by Dr. Pierre Baudoz. The testbed works in the visible and is currently using two 32x32 MEMS DMs and routinely obtains contrast better than $10^{-8}$ on a 360° dark-hole from 6 to 12 λ/D, with a clear aperture (Baudoz et al. 2018, Fig. D3). **THD2 offers to test new coronagraph designs at a high-contrast level, mostly to obtain performance in large bandwidths**. Recent work with a six-level phase mask coronagraph (Patru et al. 2018) and a dual zone phase mask coronagraph obtained a contrast of $4.10^{-8}$ between 7 and 17 λ/D in 40% bandwidth (Delorme et al. 2016, Fig D1-2).

**The THD2 team explores new WFS&C methods in the context of coronagraphy** including (i) coronagraphic focal-plane wavefront estimation for exoplanet detection (COFFEE, Paul et al. 2014), (ii) non-linear dark-hole correction (NLDH, Herscovici-Schiller et al. 2018) and (iii) the self-coherent camera (SCC, Baudoz et al. 2006). These algorithms are alternatives, each with their specific advantages, to the pair-wise probing technique used on all other testbeds (Sec 3.2.). The current goal of the testbed is the comparison of these techniques with a unique testbed and coronagraph first in a narrow bandwidth (Potier et al, in prep), and then on larger bandwidths. Finally, THD2 will reproduce these results with non-segmented on-axis apertures.

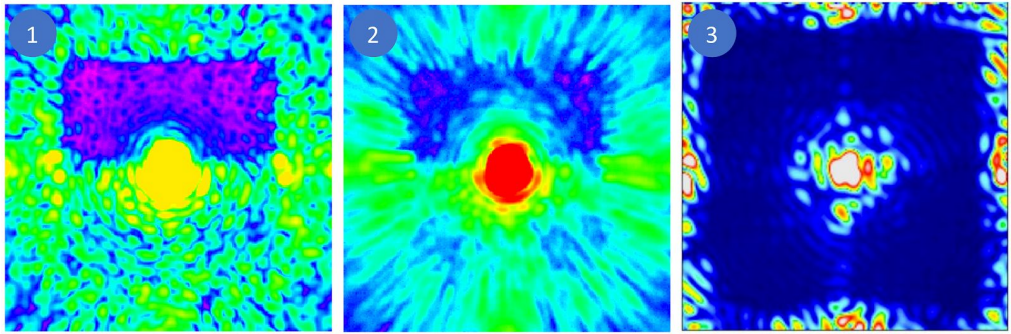

**Fig D:** *THD2 results. Single DM half-dark-hole with a 5% (1) and 50% (2) bandwidths (Delorme et al. 2016). (3) Two DMs dark-hole up to 12 λ/D (Baudoz et al. 2018).*

### 4.2. Vacuum testbeds: High-contrast imaging testbed (HCIT, JPL)

The High-Contrast Imaging Testbed (HCIT) facility at JPL (Pasadena, CA) is a large optical laboratory that hosts three optical testbeds inside vacuum chambers designed to advance coronagraph technologies for space telescopes (WFIRST, LUVOIR and HabEx): the Occulting Mask Coronagraph (OMC), the General Purpose Coronagraph Testbed (GPCT) and the Decadal Survey Testbed (DST). A decade ago, Trauger and Traub (2007) achieved the first experimental demonstration of a raw contrast levels required to image rocky exoplanets on HCIT: a contrast level of $6.10^{-10}$ over a bandwidth of 2% between 4 and 10 λ/D (180° dark-hole). Using a similar coronagraph and dark-hole shape, larger bandwidth results were obtained $3.10^{-10}$ with Δλ/λ = 2%, $6.10^{-10}$ with Δλ/λ = 10%, and $2\times10^{-9}$ with Δλ/λ = 20% (Trauger et al. 2012).

#### 4.2.1. Occulting Mask Coronagraph (OMC) testbed for WFIRST/CGI

OMC is dedicated to WFIRST/CGI technology development, using its complex discontinuous aperture (Fig. A3). **WFIRST/CGI will demonstrate, in space, two-DM active WFS&C technology for direct imaging of exoplanets, which will be crucial for HabEx/LUVOIR** **development.** OMC is used to demonstrate each of the three CGI coronagraphs modes and supports the injection and rejection of relevant disturbances in order to simulate a dynamical space environment. This testbed has already reached a number of milestones of high relevance to future mission concepts and their coronagraphic needs:

(i) low-order WFS&C of dynamic errors under flight photon flux levels (Shi et al. 2018)

(ii) dual DM broad-band starlight suppression, with a partially obscured pupil, reaching contrast levels of $5.10^{-9}$ over 18% bandwidth in a "bow-tie" dark-hole region with working angles between 3 and 9 $\lambda/D$ (Fig. E1, Cady et al. 2017 and unpublished results)

(iii) dual DM broad-band starlight suppression, with a partially obscured pupil, reaching contrast levels of $1.10^{-9}$ over 10% bandwidth in a 360 degrees dark-hole with working angles between 3 and 9 $\lambda/D$ (Fig. E2, Seo et al.2017 and unpublished results).

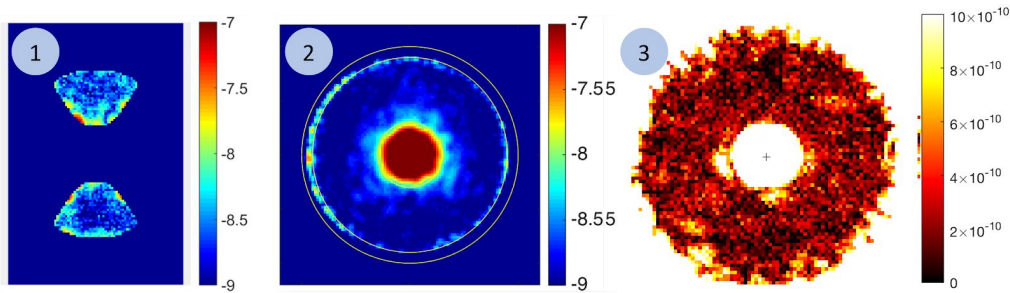

**Fig E**: *High-contrast performance obtained in the HCIT. OMC dark-holes obtained with WFIRST aperture at 10% bandwidth (1) with the SPC (contrast of $4.10^{-9}$) and (2) with the HLC ($1.10^{-9}$ contrast). (3) Phase 1a (clear aperture) Dark-hole on the DST ($4.10^{-10}$ contrast).*

### 4.2.2. General Purpose Coronagraph Testbed (GPCT)

GPCT was constructed in 2017 for testing innovative coronagraph technologies used in the HabEx and LUVOIR concepts, as well as DM technology. This testbed is used for preliminary tests for new coronagraphs (PIAACMC and Vortex) to reduce demand on the Decadal Survey Testbed for the TDEM funded tasks described in Sec. 3.1. The PIAACMC testbed is currently being commissioned at GPCT and will demonstrate a $10^{-9}$ raw contrast performance with a spectral bandwidth of 10% and an aggressive IWA of 2 $\lambda/D$ (Belikov et al. 2018a) with segmented and centrally-obstructed LUVOIR-A aperture (Fig A4-5).

### 4.2.3. The Decadal Survey Testbed (DST)

The Decadal Survey Testbed (DST) is a state-of-the-art vacuum testbed with high thermal and vibrational stability. Its nominal goal for commissioning is a raw contrast of $10^{-10}$ in an annular dark-hole of 3-10 $\lambda/D$ with a spectral bandwidth of $\Delta\lambda/\lambda \geq 10\%$. **Development will be divided into three phases that will ultimately allow the comparison of designs / WFS&C techniques in space conditions on the same testbed in a range of situations from ideal (stabilized, clear aperture) to a complete off-axis telescope simulator (segmentation, dynamic aberrations).**

**DST Phase 1** was just completed with the achievement of a raw contrast performance of $4.10^{-10}$ (Fig. E3, 360° dark-hole) with $\Delta\lambda/\lambda = 10\%$ and a clear aperture (HabEx case). For this phase, the DST used two Xinetics DMs and a classical Lyot coronagraph (no apodization). The team is currently testing **(phase 1b)** the performance in a very similar testbed layout but using MEMS DMs instead of the Xinetics DMS. The goal of this phase is to quantify tradeoffs between DM technologies for contrast performance and stability (see Sec 3.2.). This will inform the WFIRST mission and future investigators regarding DM performance trades.

In **DST Phase 2**, the team will replace the clear aperture with a pupil simulating an off-axis segmented aperture. The goal is to demonstrate <$10^{-9}$ raw contrast levels in a 360° dark-hole, 3-10 $\lambda/D$, and a spectral bandwidth of $\Delta\lambda/\lambda = 10\%$. It aims to demonstrate that a coronagraph in combination with two DMs can achieve similar performance on a clear aperture and on phased segmented aperture just by changing the shape on the DMs, provided gaps are sufficiently small.

Finally, in **DST Phase 3**, the team will test the performance with a realistic simulator of an off-axis segmented space telescope (LUVOIR B). A Zernike wavefront sensor (N'Diaye et al. 2016) will be introduced to measure dynamic aberrations and improve dark-hole stability. Finally (**phase 3b**), wavefront errors representative of segment-to-segment piston and tip/tilt errors will be simulated on the DST. The goal is to demonstrate that the Zernike Wavefront sensor can measure and control in close loop phase errors due to segment misalignments.

**After 2021**, new experiments will be designed to extend the parameter space, including on-axis apertures, larger bandwidths and dark-holes, and faster/more stable correction at the $10^{-10}$ level.

## 5. Schedule of the testbeds described in this whitepaper

| Testbeds | 2019 | 2020 | 2021 |
|---|---|---|---|
| ACE | TDEM Multi-Star — MSWC-0 (Sub-Nyquist) | TDEM Multi-Star — MSWC-s (Super-Nyquist) | |
| HCST | TDEM vortex — Clear aperture; TDEM vortex — Off-axis segmented (in phase) aperture; Fiber injection Unit | TDEM Vortex — Off-axis segmented (in phase) aperture in large bandwidth; Fiber injection Unit | |
| HiCAT | Clear aperture; TDEM APLC — On-axis segmented (in phase) | TDEM APLC — On-axis segmented w/ static segment misalignments | TDEM APLC — On-axis segmented w/ dynamic segment misalignments |
| THD2 | Comparison of sensing & control algo.; Sensing & control algo. (large bandwidth) | Sensing & control algorithms on-axis non-segmented aperture | Sensing & control algorithms on-axis non-segmented aperture |
| HCIT / GPCT | TDEM vortex — Clear ap.; TDEM PIAACMC — On-axis segmented (in phase) ap. | TDEM PIAACMC — On-axis segmented aperture in large bandwidth | |
| HCIT / DST | DST phase Ia — Clear ap.; DST phase Ib — w/ MEMS DMs; DST phase II — Off-axis segmented (in phase); TDEM Lyot — Segmented (in phase); TDEM vortex — Segmented (in phase) | | TDEM Lyot; DST phase III — Off-axis segmented w/ dynamic aberrations |

# Testbed Publications

## ACE testbed

# 8. Co-signers

| | |
|---|---|
| **California Institute of Technology**<br>Pasadena, California | Jacques-Robert Delorme, Daniel Echeverri, Nemanja Jovanovic |
| **Jet Propulsion Laboratory, California Institute of Technology**<br>Pasadena, California | Vanessa Bailey, Eric Cady, Pin Chen, Brian Kern, John Krist, Camilo Mejia Prada, Dwight Moody, Keith Patterson, A. J. Eldorado Riggs, Byoung-Joon Seo, J. Kent Wallace |
| **Laboratoire d'Études Spatiales et d'Instrumentation en Astrophysique**<br>Observatoire de Paris<br>Meudon, France | Olivier Dupuis, Axel Potier, Garima Singh, Simone Thijs |
| **Leiden University**<br>Leiden, Netherlands | Emiel Por, Matthew Kenworthy, Kelsey Miller |
| **NASA Ames**<br>Moffett Field, California | Eduardo Bendek, Eugene Pluzhnik |
| **NASA's Goddard Space Flight Center**<br>Greenbelt, Maryland | Matt Bolcar, Michael McElwain, Hari Subedi |
| **Princeton University**<br>Princeton New Jersey | He Sun, Chuanfei Dong |
| **Space Telescope Science Institute**<br>Baltimore, Maryland | Greg Brady, Keira Brooks, Tom Comeau, Jules Fowler, Rob Gontrum, Iva Laginja, Chris Moriarty, Pete Petrone, Anand Sivaramakrishnan, Kathryn St.Laurent |

*Acknowledgement: Johan Mazoyer acknowledges support for this work was provided by NASA through the NASA Hubble Fellowship grant #HST-HF2-51414 awarded by the Space Telescope Science Institute, which is operated by the Association of Universities for Research in Astronomy, Inc., for NASA, under contract NAS5-26555. The research was carried out at the Jet Propulsion Laboratory, California Institute of Technology.*